\documentclass[aps,prl,reprint,showpacs,superscriptaddress,longbibliography]{revtex4-2}
\usepackage{graphicx} 
\usepackage{amsmath}
\usepackage{graphicx,epstopdf}
\usepackage{gensymb}
\usepackage{multibib}
\newcommand{\be}{\begin{equation}}
\newcommand{\ee}{\end{equation}}
\newcommand{\bea}{\begin{eqnarray}}
\newcommand{\eea}{\end{eqnarray}}
\newcommand{\bse}{\begin{subequations}}
\newcommand{\ese}{\end{subequations}}

\usepackage{color}
\usepackage[colorlinks,bookmarks=false,citecolor=darkblue,linkcolor=red,urlcolor=blue]{hyperref}

\definecolor{darkred}{rgb}{0.7,0.0,0.0}

\definecolor{darkblue}{rgb}{0,0.02,0.45}

\definecolor{darkgreen}{rgb}{0.02,0.45,0.0}

\definecolor{violet}{rgb}{0.8,0.2,0.6}

\begin{document}

\title{Repulsive Tomonaga-Luttinger Liquid in Quasi-one-dimensional Alternating Spin-$1/2$ Antiferromagnet NaVOPO$_4$}

\author{S. S. Islam}
\thanks{These authors contributed equally to this work.}
\affiliation{School of Physics, Indian Institute of Science Education and Research, Thiruvananthapuram-695551, India}
\author{Prashanta K. Mukharjee}
\thanks{These authors contributed equally to this work.}
\affiliation{School of Physics, Indian Institute of Science Education and Research, Thiruvananthapuram-695551, India}
\author{P. K. Biswas}
\author{Mark Telling}
\affiliation{ISIS Pulsed Neutron and Muon Source, STFC Rutherford Appleton Laboratory, Harwell Campus, Didcot, Oxfordshire OX11 0QX, United Kingdom}
\author{Y. Skourski}
\affiliation{Dresden High Magnetic Field Laboratory (HLD-EMFL), Helmholtz-Zentrum Dresden-Rossendorf, 01328 Dresden, Germany}
\author{K. M. Ranjith}
\author{M. Baenitz}
\affiliation{Max Planck Institut fur Chemische Physik fester Stoffe, Nothnitzer Strasse 40, 01187 Dresden, Germany}
\author{Y. Inagaki}
\affiliation{Department of Applied Quantum Physics, Faculty of Engineering, Kyushu University, Fukuoka 819-0395, Japan}
\affiliation{Ames Laboratory, U.S. Department of Energy, Iowa State University, Ames, Iowa 50011, USA}
\author{Y. Furukawa}
\affiliation{Ames Laboratory, U.S. Department of Energy, Iowa State University, Ames, Iowa 50011, USA}
\author{A. A. Tsirlin}
\affiliation{Felix Bloch Institute for Solid-State Physics, Leipzig University, 04103 Leipzig, Germany}
\author{R. Nath}
\email{rnath@iisertvm.ac.in}
\affiliation{School of Physics, Indian Institute of Science Education and Research, Thiruvananthapuram-695551, India}
\date{\today}

\begin{abstract}
We probe the magnetic field-induced Tomonaga-Luttinger liquid (TLL) state in the bond-alternating spin-$1/2$ antiferromagnetic (AFM) chain compound NaVOPO$_4$ using thermodynamic as well as local $\mu$SR and $^{31}$P NMR probes down to milli-K temperatures in magnetic fields up to 14~T.
The $\mu$SR and NMR relaxation rates in the gapless TLL regime decay slowly following characteristic power-law behaviour, enabling us to directly determine the interaction parameter $K$ as a function of the magnetic field. These estimates are cross-checked using magnetization and specific heat data.
The field-dependent $K$ lies in the range of $0.4 < K < 1$ and indicates repulsive nature of interactions between the spinless fermions, in line with the theoretical predictions. This renders NaVOPO$_4$ the first experimental realization of TLL with repulsive fermionic interactions in hitherto studied $S=1/2$ bond-alternating AFM-AFM chain compounds.
\end{abstract}

\maketitle
Quantum phase transitions and quantum critical states in low-dimensional spin systems are an active research field in current condensed-matter physics~\cite{Sachdev2007}. Weakly coupled quasi-one-dimensional (1D) gapped spin-$1/2$ antiferromagnets (AFM) 
offer an excellent playground to study such quantum effects~\cite{Giamarchi2003,Bouillot054407,Sachdev258,Orignac140403,Meda057205}.
These gapped spin systems are highly vulnerable to the external magnetic field ($H$). Above the critical magnetic field ($H_{\rm C1}$) of gap closing, a three-dimensional (3D) magnetic long-range-order (LRO) sets in, known as Bose-Einstein-condensation (BEC) of spin-1 triplons~\cite{Giamarchi2008,Mukhopadhyay177206,Mukharjee224403}.
Further, above the 3D LRO, the gapless excitations can be universally described by a non-Fermi liquid type Tomonaga-Luttinger-liquid (TLL) theory~\cite{Tomonaga544,*Luttinger1154} of interacting spinless fermions in 1D~\cite{Giamarchi2003,Meda057205}.
The low energy excitation of the TLL state is described by two parameters: the renormalized velocity of excitations $u$ and the dimensionless interaction parameter $K$ that determines the nature and strength of the interactions between the spinless fermions~\cite{Giamarchi2003}.
The TLL model has been successfully applied to a multitude of systems such as quantum nanowires~\cite{Auslaender825}, edge state of quantum Hall systems~\cite{Grayson1062}, carbon nanotubes~\cite{Bockrath598}, and most importantly magnetic insulators~\cite{Lake329}.

Although the physics of these diverse 1D systems has been explored in the framework of TLL model, the quantitative description of TLL parameters was missing until a long time. In this regard, recent notable achievements are the experimental observation of the TLL state with attractive (i.e. $K>1$) and repulsive (i.e. $K<1$) interactions in strong-leg (C$_7$H$_{10}$N$_2$)$_2$CuBr$_4$ (DIMPY)~\cite{Schmidiger167201,Jeong106404,Jeong106402} and strong-rung (C$_5$H$_{12}$N$_2$)$_2$CuBr$_4$ (BPCB)~\cite{Klanjsek137207,Ruegg247202,Mukhopadhyay177206,Jeong106402} two-leg spin-ladders, respectively. Moreover, a remarkable control of the TLL properties by means of field tuning of the $u$ and $K$ parameters has also been demonstrated in these weakly coupled spin-ladder systems~\cite{Klanjsek137207,Jeong106404,Jeong106402}. Here, $H$ acts as a chemical potential that controls the filling of the fermion band and, therefore, tunes the dynamics of the TLL state. \textbf{TLL with repulsive interaction is also demonstrated in spin-$1/2$ uniform Heisenberg/Ising spin chains~\cite{Lake329,Klanjsek057205,Kono037202,Halg104416,Horvatic220406}.}
\begin{figure*}
	\includegraphics [width = \linewidth]{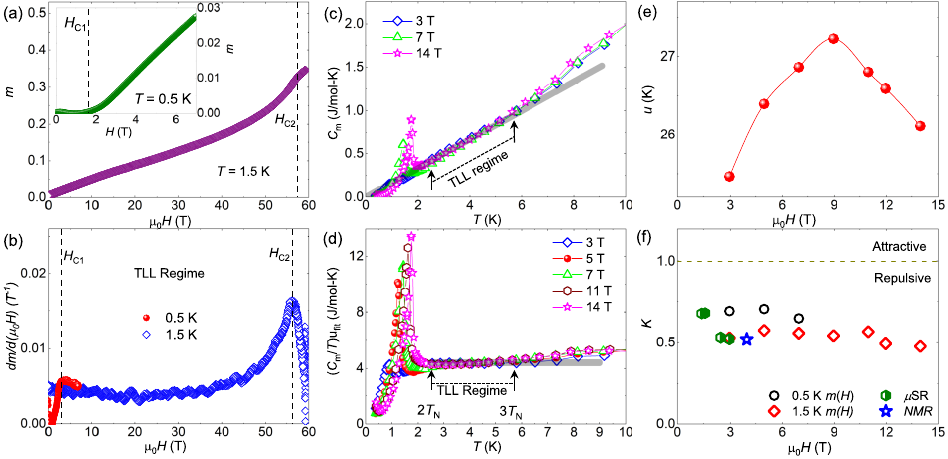}
	\caption{(a) $m$ vs $H$ measured at $T = 1.5$~K. Inset: $m$ (after impurity subtraction) vs $H$ at $T=0.5$~K. (b) $dm/dH$ vs $H$ together for $T = 1.5$~K and 0.5~K. (c) Magnetic specific heat ($C_{\rm m}$) vs $T$ for three selected fields. Dashed line guided by two upward arrows marks the $T$-linear TLL regime. (d) Normalized $C_{\rm m}$ vs $T$ in different fields above $H_{C1}$. The linear TLL regime is marked by two arrows. (e) Normalized velocity $u$ vs $H$, estimated from the $C_{\rm m}$ analysis. (f) TLL exponent $K$ vs $H$ deduced from magnetization, specific heat, $\mu$SR, and NMR $1/T_1$ experiments.}
	\label{Fig1}
\end{figure*}

Apart from spin-$1/2$ uniform Heisenberg/Ising spin chains and spin ladders, the Heisenberg bond-alternating spin-1/2 chains have been predicted to evince field-induced TLL state of interacting spinless fermions~\cite{Abouie184437,Sakai251}. The nature of interactions between these fermions, attractive or repulsive, should be determined by the sign and hierarchy of the nearest-neighbor exchange couplings. One expects attractive interaction for ferromagnetic (FM)-AFM spin chains and repulsive interaction for AFM-AFM spin chains in the TLL state~\cite{Sakai251}. On the experimental side, only a very few bond-alternating spin-$1/2$ chain systems are reported where TLL behavior is anticipated from the thermodynamic measurements. They include copper nitrate Cu(NO$_3$)$_2$·2.5D$_2$O~\cite{Willenberg060407} and pentafluorophenyl nitronyl nitroxide F$_5$PNN~\cite{Matsushita020408} with AFM-AFM spin-chains, zinc-verdazyl complex~\cite{Yamaguchi085117} with FM-AFM spin-chains etc. However, none of these reports present a detailed quantitative analysis and verification of the nature of interactions (attractive/repulsive) in the TLL state.

The recent discovery of an alternating spin-1/2 chain compound NaVOPO$_4$~\cite{Mukharjee144433} appears to be promising in this context. It features quasi-1D bond-alternating spin-$1/2$ AFM-AFM chains formed by VO$_6$ octahedra with leading exchange coupling $J/k_{\rm B} \simeq 39$~K along the chain direction with a weak alternation ratio $\alpha (=J'/J) \simeq 0.98$. The ground state is a singlet with a tiny zero-field spin-gap $\Delta_{0}/k_{\rm B} \simeq 2$~K that corresponds to a small accessible critical field of $\mu_{0}H_{\rm C1}\simeq 1.6$~T. The critical field of saturation is $H_{\rm C2} \simeq 57$~T. An external field $H \geq H_{\rm C1}$ closes the gap and gives rise to a 3D LRO ($T_{\rm N}$) which is described well by 3D BEC of triplons. Since bond-alternating spin chains are predicted to show evidence of both TLL and BEC phases, NaVOPO$_4$ gives a unique opportunity to study the crossover effect from 1D TLL at high-$T$s to 3D BEC at low-$T$s. Moreover, the low value of $H_{\rm C1}$ provides an added advantage to explore the critical TLL behavior under laboratory fields.

In this letter, we present a detailed study of the TLL state in NaVOPO$_4$. Using magnetization and specific heat and also powerful local tools such as nuclear magnetic resonance (NMR) and muon-spin relaxation ($\mu$SR), we demonstrate that the interaction between spinless fermions is repulsive in nature within the TLL state, which is in accordance with the theoretical predictions~\cite{Sakai251}. Moreover, from the temperature-dependent NMR spin-lattice relaxation rate [$1/T_{\rm 1}$] we estimate the exact value of the TLL parameter ($K$) employing the recent analytical results reported by Dupont $et~al$~\cite{Dupont094403,Horvatic220406}. Our study not only links the 1D TLL theory to bond-alternating AFM-AFM Heisenberg spin-1/2 chain systems but also provides insight about the nature and strength of interactions between the spinless fermions in the gapless TLL state.

All the experiments were performed on the polycrystalline NaVOPO$_4$ sample. Details of the magnetization, specific heat, and $^{31}$P NMR measurements are described in Ref.~[\onlinecite{Mukharjee144433}].
$\mu$SR measurements were carried out in zero-field (ZF) as well as in longitudinal-fields (LFs) down to 50~mK [see Supplemental Material (SM)~\cite{supplementary} for details].


We begin the discussion of prospective TLL behavior in NaVOPO$_4$ from the magnetization and specific heat. Figure~\ref{Fig1}(a) presents the reduced magnetization ($m$) vs $H$ measured at $T = 1.5$~K, whereas the inset shows the corrected $m$ vs $H$ curve obtained after subtraction of the extrinsic paramagnetic contribution (see Ref.~\cite{Mukharjee144433}). Here, the dimensionless quantity $m$ is nothing but the expectation value of spin $\left\langle {S}\right\rangle$ [i.e. $m = \frac {M}{g\mu_{\rm B}} = \left\langle {S}\right\rangle$], calculated normalizing the measured magnetization by the Land\'{e} $g$-factor $g \simeq 1.951$ (obtained from ESR experiments~\cite{Mukharjee144433}). As expected for a gapped quantum magnet, at $T = 0.5$~K $m$ remains nearly zero up to $H_{\rm C1}$ and $dm/dH$ exhibits a rounded peak at $H_{\rm C1} \simeq 1.6$~T as shown in Fig.~\ref{Fig1}(b). Similarly, at $T=1.5$~K, $m$ saturates in an applied field of $H_{\rm C2}\simeq 57$~T where $dm/dH$ shows another rounded maximum. These two rounded peaks in $dm/dH$ [Fig.~1(b)] indicate the crossover in and out of the 1D TLL regime, respectively~\cite{Willenberg060407,Jeong106402,Wang5399}.

The magnetic specific heat ($C_{\rm m}$) as a function of temperature is depicted in Fig.~\ref{Fig1}(c) at three representative fields (for $H > H_{\rm C1}$)~\footnote{$C_{\rm m}$ was obtained by subtracting the phonon contribution from the total specific heat. In order to estimate the phonon part, we fitted the high-temperature part of the total specific heat in zero field by a sum of multiple Debye terms as in Ref.~\cite{Mukharjee144433} and the fit was extrapolated down to low temperatures. The same phonon contribution was also used to estimate $C_{\rm m}$ in different fields.}. A striking feature observed in $C_{\rm m} (T)$ is the appearance of a $T$-linear regime (2.5~K to 7~K) above $T_{\rm N}$ which remains almost independent of field. Such a behaviour is a robust signature of the TLL state~\cite{Hagiwara147203}. In the TLL critical regime, the molar magnetic specific heat ($C_{\rm m}$) is related to $u$ as~\cite{Giamarchi2003,Jeong106402,Galeski237201}
\begin{equation}
	\label{C_TLL}
	C_m = N_{\rm A}\frac{\pi k_{\rm B}T}{6u},
\end{equation}
where $N_{\rm A}$ is the Avogadro's constant. To estimate $u$, we fitted $C_{\rm m}(T)$ in the $T$-linear region by Eq.~\eqref{C_TLL} and the corresponding $u$ values obtained for different fields are plotted as a function of $H$ in Fig.~\ref{Fig1}(e). In order to check the reliability of the estimation of $u$, $C_{\rm m}$ at different fields is normalized with respect to $u$  [i.e. $(C_{\rm m}/T)u_{\rm fit}$] and plotted as a function of $T$ in Fig.~\ref{Fig1}(d). Interestingly, the $(C_{\rm m}/T)u_{\rm fit}$ curves above 2.5~K collapse on a flat line irrespective of the field value, which is a clear demonstration of the TLL regime~\cite{Jeong106402}. Further, with increasing field, the $T$-independent regime is widened, an indication of the stabilization of the TLL state in higher fields~\cite{Hagiwara147203}. The variation of $u$ in Fig.~\ref{Fig1}(e) is found to be reminiscent of the behavior reported for the spin-ladder compound BPCB with repulsive TLL interactions~\cite{Jeong106402}.

Another crucial parameter that characterizes TLL correlations is $K$. The relation between $K$ and $u$ can be expressed as~\cite{Jeong106402}
\begin{equation}
\frac{\partial{m}}{\partial{(\mu_{0}H)}} = \frac{g \mu_{\rm B}}{2\pi k_{\rm B}} \left(\frac{K}{u}\right).
	\label{relation btw Ku}
\end{equation}
The $u$ values obtained from the specific heat analysis and differential magnetization $\partial{m}/\partial{(\mu_{0}H)}$ data at 0.5~K and at 1.5~K from Fig.~\ref{Fig1}(b) are used in Eq.~\eqref{relation btw Ku} to determine $K$. The obtained $K$ values plotted in Fig.~\ref{Fig1}(f) are in the range of $0.5 < K < 1$. The value of $K$ less than 1 indicates repulsive TLL interactions. Moreover, the value of $K$ is also found to be more than 0.5, which suggests Heisenberg-type interactions with dominant transverse AFM fluctuations~\cite{Horvatic220406}. However, we note that the above method of estimating the TLL parameters may be biased by the quality of the experimental data~\cite{Jeong106402}. Therefore, we back our conclusions using local probes like $\mu$SR and NMR, both of which give direct access to the low-energy spin dynamics.

\begin{figure*}
	\includegraphics[width = \linewidth]{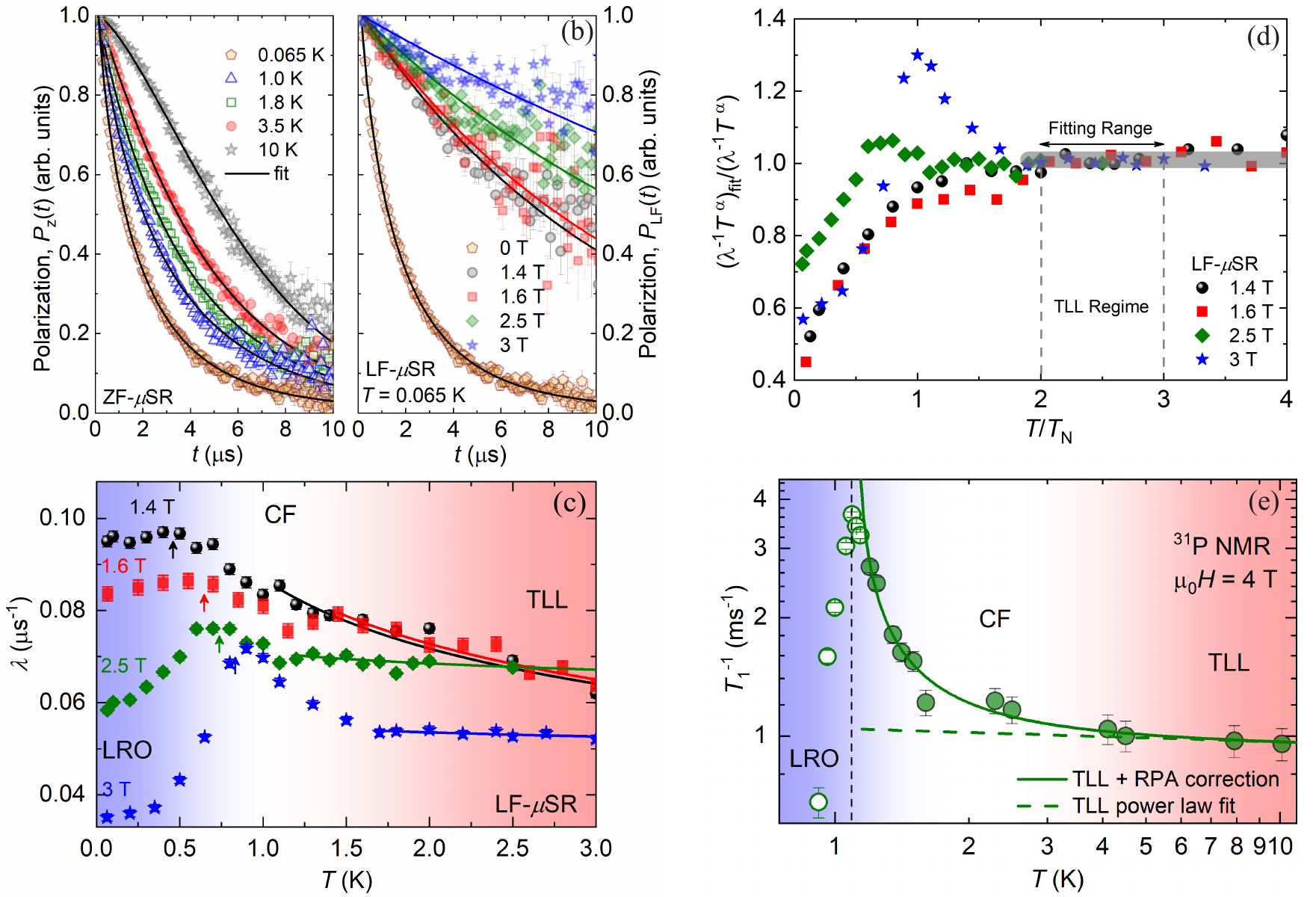}
	\caption{\label{Fig2} (a) ZF-$\mu$SR polarization measured at different temperatures down to 65~mK. The solid lines are the fits using the stretched exponential function, described in SM~\cite{supplementary}. (b) $\mu$SR polarization at 65~mK and in different LFs. The solid lines are the fits using single exponential function, described in the text. (c) Temperature dependent $\mu$SR depolarization rate ($\lambda$) in different LFs. The solid lines are fits using Eq.~\eqref{TLL_muSR}. (d) Normalized $\lambda ^{-1} T^\alpha$ vs $T/T_{\rm N}$ to show the collapse of data sets in different fields onto one line. (e) The RPA+TLL fit (solid line) using Eq.~\eqref{RPA+TLL} to the $1/T_1(T)$ data at 4~T. The filled circles represent the data set considered for the fitting. The horizontal dashed line indicates the pure TLL contribution to the fit. The vertical dashed line denotes the $T_{\rm N}$ determined by the fit.}
\end{figure*}
Figure~\ref{Fig2}(a) shows representative muon spin polarization curves measured at different temperatures from 10~K to 65~mK in zero field. At 10~K, it shows a Gaussian-like behavior. As the temperature is lowered, the shape of $P_z(T)$ eventually changes to exponential-like behavior.
It exhibits neither coherent oscillations nor a loss of initial polarization down to 65~mK, ruling out the transition to a magnetic LRO state. The spectra also do not decay to $1/3$ of the initial polarization 
at long times, characteristic of a quasi-static local magnetic field at the muon stopping site, thus excluding a spin-glass type transition~\cite{Uemura546,Uemura3306}. 
These observations are consistent with a singlet/disordered ground state reported previously in zero field~\cite{Mukharjee144433}. However, a weak increase in the muon depolarization rate ($\lambda$) below the temperature scale of $\Delta_0/k_{\rm B} \simeq 2$~K can be ascribed to the quasi-free spins that arise from singlets broken by implanted muon or from a small amount of impurity/orphan spins (see SM~\cite{supplementary}).

To look for the temperature-dependent electronic spin dynamics in NaVOPO$_4$ at higher fields ($ H \geq  H_{\rm C1}$), we have measured $\mu$SR response in four different LFs (i.e. 1.4, 1.6, 2.5 and 3~T). Magnetic field at the muon stop sites includes contributions from nuclear as well as electronic field fluctuations. In an applied longitudinal field, the nuclear contribution to the muon spin relaxation is expected to be fully quenched and thus one can reliably probe the electronic spin dynamics. Figure~\ref{Fig2}(b) presents the muon spin polarization at the lowest measured temperature of 65~mK in different LFs. The increasing slope of the curves with the applied field suggests a slowing down of electronic fluctuations with increasing field, in the field-induced ordered state~\cite{Mukharjee144433}.
Remarkably, the LF-$\mu$SR polarization curve at the highest measured LF of $3$~T also exhibits substantial relaxation, which is not completely quenched, implying electronic nature of the fluctuations.
As the compound shows the onset of field-induced magnetic LRO above $H_{\rm C1}$, one would expect distinct oscillations in the muon spin polarization below $T_{\rm N}$ or at least at the lowest measured temperature 65~mK. However, in our case we did not observe any such oscillations which could be due to the broad distribution of the local magnetic field at the muon site~\cite{Xu064425}. Similar muon response without visible oscillations has been observed in the ordered state of other gapped quantum AFMs~\cite{Lancaster207201,Moller020402}.

The muon spin polarization curves over the entire measured temperature range and for all LFs follow a simple exponential relaxation [Fig.~\ref{Fig2}(b)],
\begin{equation}
\label{muSR_eqtn}
P_{\rm LF}(t) = P_{0}e^{-\lambda t}+P_{b},
\end{equation}
where $P_{0}$ is the initial muon spin polarization and $P_{b}$ is the non-relaxing background component arising from the sample holder or the sample background. For each field, both $P_0$ and $P_b$ are nearly temperature-independent.
The estimated $\lambda$ is plotted against temperature in Fig.~\ref{Fig2}(c) where we can identify three distinct temperature regimes. In the low-temperature regime, as the temperature rises, $\lambda$ increases and shows an anomaly at $T_{\rm N}$, suggesting the onset of magnetic LRO. With increasing LF, the anomaly at $T_N$ becomes more pronounced and shifts towards higher temperatures. The peak position in different fields traces the phase boundary (magnon BEC) shown in Ref.~[\onlinecite{Mukharjee144433}]. The region immediately above $T_{\rm N}$ is the critical fluctuation (CF) regime where thermal fluctuations dominate the relaxation. In the CF region, $\lambda$ shows a decreasing behavior until the power-law behavior sets in above $\sim 2T_{\rm N}$ indicating TLL fluctuations. We note in passing that a peak in $\lambda$ is observed even at $H = 1.4$~T, reflecting that the actual $H_{\rm C1}$ could be lower than 1.4~T.

In quantum spin chains, the spin-spin correlation decays following the power law, with the exponent being a simple function of $K$~\cite{Stolze4319}. Hence, one can estimate the TLL parameter ($K$) by modeling the $\mu$SR depolarization rate in the gapless TLL regime (above $T_{\rm N}$) with a power law of the form~\cite{Moller020402}
\begin{equation}
	\label{TLL_muSR}
	\lambda = T_{\rm TLL}^{-1}(T) \propto T^{1/2K-1}.
\end{equation}
We fitted the data in Fig.~\ref{Fig2}(c) by Eq.~\eqref{TLL_muSR} in the temperature window $2T_N \leq T \leq 3T_N$. To validate the choice of this temperature window, we normalized $\lambda$ as $\lambda/T^\alpha$ (where $\alpha = 1/2K-1$) and plotted it against the reduced temperature $T/T_N$ [see Fig.~\ref{Fig2}(d)]. Clearly, the data for all the four fields collapse on a straight line for $2T_N \leq T \leq 3T_N$ and also beyond $3T_N$, highlighting the universal power-law behavior in the TLL regime and confirming the reliability of the estimated $\alpha$ or $K$.
Further, to ensure that the extracted $K$ value is not affected by a particular choice of the fitting range, we fitted the data four times~\cite{Moller020402}: first over the full fitting window $2T_N \leq T \leq 3T_N$, then excluding either of the end points, and finally excluding both the end points. Consequently, the final value of $K$ is taken as the inverse-variance weighted average and is plotted against $H$ in Fig.~\ref{Fig1}(f). Thus, the obtained TLL parameter $K <1$ once again proves the repulsive interactions between the spinless fermions in the TLL state. Moreover, the $K$ value converges towards 1 as the field approaches $H_{\rm C1}$, suggesting that the system approaches the non-interacting and/or 1D quantum critical regime, consistent with the theory~\cite{Jeong106404}.

We note that a simple power-law dependence of the relaxation rate, which is predicted for a purely 1D spin system, underestimates the value of $K$ since it does not take into account the 3D critical fluctuations that  are predominant at low temperatures, especially in the region close to $T_{\rm N}$. In the past, attempts have been made to estimate $K$ for the spin-ladder compound DIMPY from the temperature-dependent NMR $1/T_1$ using the generic power law~\cite{Jeong106404,Klanjsek057205}. 
However, the obtained values of $K$ were not in good agreement with the theory. This mismatch was ascribed to the enhancement of the relaxation due to the 3D critical fluctuations in the vicinity of $T_{\rm N}$~\cite{Dupont094403}. Recently, the effect of this fluctuation near $T_{\rm N}$ has been theoretically modeled by Dupont~$et.~al.$~\cite{Dupont094403} using a random phase approximation (RPA). Thereafter, this approximation was used by Horvati$\acute{c}$~$et.~al$.~\cite{Horvatic220406} as a correction to the TLL power-law behavior that takes the form (RPA+TLL)
\begin{equation}
1/T_1 \propto T^{1/2K-1} \phi (K, T_{\rm N}/T).
\label{RPA+TLL}
\end{equation}   
Here, $\phi(K, T_{\rm N}/T)$ is the RPA correction factor, which is a multiplicative correction function of $K$ and $T_{\rm N}$~\cite{Horvatic220406,Dupont094403}.
The validity of this method was independently tested on Ising spin-$1/2$ chain compound BaCo$_2$V$_2$O$_8$ and spin-ladder compound DIMPY~\cite{Horvatic220406}.
The fitting window in this method can be extended to $1.2 T_{\rm N}$ because the effect of critical fluctuations is included.

In Fig.~\ref{Fig2}(e), we show the temperature dependence of $1/T_{1}$ in a magnetic field of 4~T, which is well above $H_{\rm C1}$. The sharp $\lambda$-type anomaly at $T_{\rm N} \simeq 1.09$~K reflects the onset of a 3D magnetic LRO. The non-linear fit using the analytical expression [Eq.~\eqref{RPA+TLL}] returns $K \simeq 0.52$ and $T_{\rm N} \simeq 1.11$~K. This $K$ value clearly reflects that the interactions are repulsive in nature and are in line with our previous analysis. Moreover, the obtained $T_{\rm N}$ value is also in excellent agreement with our experimental data.

Quantum magnets demonstrating both BEC and TLL physics are rare. The majority of known materials feature very high critical fields or show the onset of field-induced LRO at relatively high temperatures. In this regard, NaVOPO$_4$ is a very promising candidate system in which both BEC and TLL physics can be realized due to the low critical field of $\mu_{0}H_{\rm C1} \simeq 1.4$~T. Indeed, we are able to tune the TLL parameter ($K$) in the gapless state by applying the magnetic field. For a comparison, we have plotted $K$ vs $H$ in Fig.~\ref{Fig1}(f) extracted from the specific heat, NMR, and $\mu$SR data. The $K$ values from the different experiments are consistent and remain below unity, indicating the repulsive nature of the interactions. Moreover, as we move towards the non-interacting regime, i.e., close to $H_{\rm C1}$, the $K$ value increases slowly towards unity. The overall behavior of $u$ with $H$ is symmetric and its shape is similar to the previously reported spin-ladder compound BPCB with repulsive TLL interactions~\cite{Jeong106402,Schmidiger167201,Bouillot054407}. Our results for NaVOPO$_4$ offer the first experimental observation of the TLL physics with repulsive interactions in bond-alternating spin chains. This finding is in an excellent agreement with theory~\cite{Sakai251}.

In summary, we have obtained direct experimental confirmation for repulsive interactions between spinless fermions in the TLL state of NaVOPO$_4$. Several experimental methods consistently show repulsive interactions in the gapless TLL regime above $H_{\rm C1}$ and at temperatures above $\sim 2T_{\rm N}$. The $\mu$SR and NMR relaxation rates, which are measures of the correlation time, follow power-law behavior and provide direct estimates of the interaction parameter $K$. A more exact value of $K$ is obtained from the analysis of $1/T_1$ using the recently proposed RPA+TLL method. Our results rigorously establish NaVOPO$_4$ as the first candidate bond-alternating spin-$1/2$ AFM-AFM chain system manifesting gapless TLL phase with repulsive interactions, validating the theoretical predictions.

\acknowledgments
We would like to acknowledge SERB, India for financial support bearing sanction Grant No.~CRG/2022/000997. We also acknowledge the $\mu$SR facility at STFC, ISIS (UK) for providing the beam time.
Work at the Ames Laboratory was supported by the U.S. Department of Energy, Office of Science, Basic Energy Sciences, Materials Sciences and Engineering Division. The Ames Laboratory is operated for the U.S. Department of Energy by Iowa State University under Contract No~ DEAC02-07CH11358. We also acknowledge the support of the HLD-HZDR, member of European Magnetic Field Laboratory (EMFL).

%

\end{document}